\documentclass{article}%
\usepackage{amsmath}
\usepackage{amssymb}
\usepackage{hyperref}
\usepackage{amsfonts}
\usepackage{graphicx}%
\begin{document}

\title{Lie Groups and Propagators Exemplified}
\author{T.L. Curtright$^{\S }$, Z. Cao, A. Peca, D. Sarker, and B.D. Shrestha$\medskip
$\\Department of Physics, University of Miami, Coral Gables, Florida 33124\\$^{\S }${\footnotesize curtright@miami.edu\medskip}}
\date{}
\maketitle

\begin{abstract}
Concise methods are used to compute the propagator for a non-relativistic
particle subject to a potential with $x^{2}$ and $1/x^{2}$ terms.

\end{abstract}

Time-development is a -- if not \emph{the} -- central issue in quantum
mechanics. \ The abundance of papers on the propagator for the harmonic
oscillator \cite{Coleman} bears witness to this fact, especially in this
journal \cite{Holstein}-\cite{MoriconiM}. \ Here is yet another paper on this
subject, hopefully more concise than the average and at least as useful for
purposes of pedagogy. \ The present paper differs from most of those just
cited in that it does not use the path integral formulation of quantum
mechanics, nor does it use explicit properties of the energy eigenfunctions
for the oscillator. \ The essential ingredient used here is an algebraic
identity for the special linear group in two dimensions, $SL\left(  2,%
\mathbb{C}
\right)  $. \ While some of the pedagogical papers cited also use algebraic
methods, in particular raising and lowering operators for the oscillator
\cite{Barone,Shao,MoriconiM}, none exploits Lie group methods so succinctly as
the analysis to follow. \ 

Consider the following identity for elements of the group $SL\left(  2,%
\mathbb{C}
\right)  $ that is relevant for this problem. \ The identity can be realized
in terms of canonical operators $x$ and $p$ for the quantum oscillator. \ So
expressed, the operator identity is \cite{TSvK}%
\begin{align}
&  \exp\left(  -\frac{it}{\hbar}\left(  \frac{1}{2m}\left(  p^{2}%
+\frac{\lambda}{x^{2}}\right)  +\frac{1}{2}~m\omega^{2}x^{2}\right)  \right)
\label{TSvKIdentity}\\
&  =\exp\left(  -\frac{im\omega}{2\hbar}~x^{2}\tan\left(  \omega t/2\right)
\right)  \exp\left(  -\frac{i}{2m\omega\hbar}~\left(  p^{2}+\frac{\lambda
}{x^{2}}\right)  \sin\left(  \omega t\right)  \right)  \exp\left(
-\frac{im\omega}{2\hbar}~x^{2}\tan\left(  \omega t/2\right)  \right) \nonumber
\end{align}
where as usual, $\hbar$, $m$, $\lambda$, \& $\omega$ are constants, and $t$ is
the time. \ This identity follows from the elementary commutation relations
that provide a realization of the Lie algebra $sl\left(  2,%
\mathbb{C}
\right)  $,%
\begin{equation}
\left[  x^{2},p^{2}+\frac{\lambda}{x^{2}}\right]  =2i\hbar\left(
xp+px\right)  \ ,\ \ \ \left[  x^{2},xp+px\right]  =4i\hbar\left(
x^{2}\right)  \ ,\ \ \ \left[  p^{2}+\frac{\lambda}{x^{2}},xp+px\right]
=-4i\hbar\left(  p^{2}+\frac{\lambda}{x^{2}}\right)  \ , \label{sl2}%
\end{equation}
which in turn follow from $\left[  x,p\right]  =i\hbar$. \ Upon
differentiating both left- and right-hand sides with respect to the time, with
the obvious initial condition at $t=0$, the identity (\ref{TSvKIdentity}) can
be confirmed by re-ordering the various operator expressions through the use
of the commutators (\ref{sl2}) in conjunction with the expansion
\begin{equation}
e^{G}Oe^{-G}=O+\sum_{n=1}^{\infty}\frac{1}{n!}\underbrace{\left[  G,\left[
\cdots,\left[  G,O\right]  \right]  \right]  }_{n\text{-fold commutation with
}G}%
\end{equation}
This straightforward but tedious method to establish (\ref{TSvKIdentity})\ is
left as an exercise for the conscientious reader.

But perhaps the easiest way to check the identity (\ref{TSvKIdentity}) is to
use a fundamental faithful representation of the $sl\left(  2,%
\mathbb{C}
\right)  $ algebra (\ref{sl2}) as $2\times2$ traceless matrices. \ In the
oscillator context, such a representation is given by
\begin{equation}
x^{2}\rightarrow\left(
\begin{array}
[c]{cc}%
0 & 2\hbar\\
0 & 0
\end{array}
\right)  \ ,\ \ \ p^{2}+\frac{\lambda}{x^{2}}\rightarrow\left(
\begin{array}
[c]{cc}%
0 & 0\\
2\hbar & 0
\end{array}
\right)  \ ,\ \ \ xp+px\rightarrow\left(
\begin{array}
[c]{cc}%
-2i\hbar & 0\\
0 & 2i\hbar
\end{array}
\right)  \label{2x2}%
\end{equation}
Hence the above group elements are realized as the $2\times2$ matrices.%
\begin{equation}
\exp\left(  -i\alpha x^{2}\right)  \rightarrow\left(
\begin{array}
[c]{cc}%
1 & -2i\hbar\alpha\\
0 & 1
\end{array}
\right)  \ ,\ \ \ \exp\left(  -i\beta\left(  p^{2}+\frac{\lambda}{x^{2}%
}\right)  \right)  \rightarrow\left(
\begin{array}
[c]{cc}%
1 & 0\\
-2i\hbar\beta & 1
\end{array}
\right)
\end{equation}
as well as%
\begin{equation}
\exp\left(  -i\alpha x^{2}\right)  \exp\left(  -i\beta\left(  p^{2}%
+\frac{\lambda}{x^{2}}\right)  \right)  \exp\left(  -i\alpha x^{2}\right)
\rightarrow\left(
\begin{array}
[c]{cc}%
1-4\alpha\beta\hbar^{2} & -4i\alpha\hbar\left(  1-2\alpha\beta\hbar^{2}\right)
\\
-2i\beta\hbar & 1-4\alpha\beta\hbar^{2}%
\end{array}
\right)
\end{equation}
where $\alpha=\frac{m\omega}{2\hbar}\tan\left(  \omega t/2\right)  $ and
$\beta=\frac{1}{2m\omega\hbar}\sin\left(  \omega t\right)  $. \ The last of
these group elements is exactly the same as%
\begin{gather}
\exp\left(  -\frac{it}{\hbar}\left(  \frac{1}{2m}\left(  p^{2}+\frac{\lambda
}{x^{2}}\right)  +\frac{1}{2}m\omega^{2}x^{2}\right)  \right)  \rightarrow
\exp\left(  -\frac{it}{2m\hbar}\left(
\begin{array}
[c]{cc}%
0 & 0\\
2\hbar & 0
\end{array}
\right)  -\frac{it}{2\hbar}m\omega^{2}\left(
\begin{array}
[c]{cc}%
0 & 2\hbar\\
0 & 0
\end{array}
\right)  \right) \nonumber\\
=\left(
\begin{array}
[c]{cc}%
\cos\left(  \omega t\right)  & -im\omega\sin\left(  \omega t\right) \\
-\frac{i}{m\omega}\sin\left(  \omega t\right)  & \cos\left(  \omega t\right)
\end{array}
\right)
\end{gather}
thereby establishing the identity (\ref{TSvKIdentity}). \ Note that there is
no dependence on $\lambda$ when using this realization of the group elements
as $2\times2$\ matrices \cite{Exercise}.

With the identity (\ref{TSvKIdentity}) in hand, the propagator for the
Hamiltonian
\begin{equation}
H\left(  \lambda,\omega\right)  =\frac{1}{2m}\left(  p^{2}+\frac{\lambda
}{x^{2}}\right)  +\frac{1}{2}m\omega^{2}x^{2}%
\end{equation}
immediately reduces to the propagator for the same Hamiltonian shorn of the
$x^{2}$ term,%
\begin{equation}
H_{0}\left(  \lambda\right)  =\frac{1}{2m}\left(  p^{2}+\frac{\lambda}{x^{2}%
}\right)
\end{equation}
albeit with a simple re-parameterized time, $t\rightarrow\frac{\sin\left(
\omega t\right)  }{\omega}$. \ That is to say, the identity
(\ref{TSvKIdentity}) gives a relation between localized matrix elements
$\left\langle x_{1}\right\vert \cdots\left\vert x_{2}\right\rangle $ for
complete eigenstates of the operator $x$, where $\left\langle x_{1}\right\vert
x=x_{1}\left\langle x_{1}\right\vert $ and $x\left\vert x_{2}\right\rangle
=x_{2}\left\vert x_{2}\right\rangle $, normalized such that $\int\left\vert
x_{1}\right\rangle \left\langle x_{1}\right\vert ~dx_{1}=1$. \ Namely,
\begin{align}
&  \left\langle x_{1}\right\vert \exp\left(  -iH\left(  \lambda,\omega\right)
t/\hbar\right)  \left\vert x_{2}\right\rangle \label{Element}\\
&  =\exp\left(  -\frac{im\omega}{2\hbar}~x_{1}^{2}\tan\left(  \omega
t/2\right)  \right)  ~\left\langle x_{1}\right\vert \exp\left(  -i~H_{0}%
\left(  \lambda\right)  \frac{\sin\left(  \omega t\right)  }{\hbar\omega
}\right)  \left\vert x_{2}\right\rangle ~\exp\left(  -\frac{im\omega}{2\hbar
}~x_{2}^{2}\tan\left(  \omega t/2\right)  \right) \nonumber
\end{align}
It only remains to compute the matrix element involving $H_{0}$.

The simplest case is $\lambda=0$ for which there is no $1/x^{2}$ potential.
\ In that case, the problem reduces to the familiar free particle propagator,
most easily evaluated by inserting complete sets of momentum eigenfunctions,
normalized such that $\int\left\vert p_{1}\right\rangle \left\langle
p_{1}\right\vert ~dp_{1}=1$. \ The transformation to position eigenstates is
then given by plane waves $\left\langle x_{1}|p_{1}\right\rangle =\frac
{1}{\sqrt{2\pi\hbar}}\exp\left(  ip_{1}x_{1}/\hbar\right)  $, and hence%
\begin{equation}
\left\langle x_{1}\right\vert \exp\left(  -\frac{itp^{2}}{2m\hbar}\right)
\left\vert x_{2}\right\rangle =\frac{1}{2\pi\hbar}\int dp_{1}\exp\left(
\frac{i}{\hbar}~p_{1}\left(  x_{1}-x_{2}\right)  -\frac{itp_{1}^{2}}{2m\hbar
}\right)  =\sqrt{\frac{m}{2\pi i\hbar t}}\exp\left(  \frac{im}{2\hbar
t}\left(  x_{1}-x_{2}\right)  ^{2}\right)
\end{equation}
The final exponential involves the action for a classical path connecting
$x_{1}$ and $x_{2}$ in time $t$, as is well-known. \ 

Re-parameterizing to a periodic time variable, $t\rightarrow\frac{\sin\left(
\omega t\right)  }{\omega}$, and combining this last relation with
(\ref{Element})\ then gives immediately the expected result for the harmonic
oscillator \cite{Feynman}.%
\begin{align}
&  \left\langle x_{1}\right\vert \exp\left(  -\frac{it}{\hbar}\left(  \frac
{1}{2m}~p^{2}+\frac{1}{2}~m\omega^{2}x^{2}\right)  \right)  \left\vert
x_{2}\right\rangle \nonumber\\
&  =\sqrt{\frac{m\omega}{2\pi i\hbar\sin\left(  \omega t\right)  }}%
~\exp\left(  -\frac{im\omega}{2\hbar}~x_{1}^{2}\tan\left(  \omega t/2\right)
+\frac{im\omega}{2\hbar\sin\left(  \omega t\right)  }\left(  x_{1}%
-x_{2}\right)  ^{2}-\frac{im\omega}{2\hbar}~x_{2}^{2}\tan\left(  \omega
t/2\right)  \right) \nonumber\\
&  =\sqrt{\frac{m\omega}{2\pi i\hbar\sin\left(  \omega t\right)  }}%
~\exp\left(  \frac{im\omega}{2\hbar}\left(  x_{1}^{2}+x_{2}^{2}\right)
\cot\left(  \omega t\right)  -\frac{im\omega}{\hbar\sin\left(  \omega
t\right)  }~x_{1}x_{2}\right)  \label{SHO}%
\end{align}
where $\cot\left(  2\theta\right)  =-\tan\left(  \theta\right)  +1/\sin\left(
2\theta\right)  $ was used in the last step. \ Once again, the final
exponential involves the action for a classical path connecting $x_{1}$ and
$x_{2}$ in time $t$, as is well-known.

The situation with $\lambda\neq0$ is more challenging, in part because of
singular behavior if the potential is too attractive when $\lambda<0$, as will
become apparent in the final result for the propagator to be given below.
\ For this and other reasons, it is convenient to define
\begin{equation}
\lambda=\hbar^{2}\left(  n^{2}-1/4\right)
\end{equation}
where $n$ is a positive, dimensionless number. \ Clearly, for $\lambda>0$,
i.e. $n>1/2$, the potential is repulsive so that a classical particle would be
confined to the half-line with $x>0$, and would only be allowed to have
positive energy. \ This is also true for the quantized system if $\lambda>0$,
and this will be assumed to be the case in the following. \ However, the final
results will turn out to be valid for $\lambda\geq-\hbar^{2}/4$, i.e. $n\geq
0$. \ 

The propagator for $H_{0}$ with $\lambda>0$ can be obtained by a method
similar to the plane wave expansion used for $\lambda=0$, only now the
relevant set of states involves Bessel functions, $J_{n}\left(  kx\right)  $
\cite{Watson}. \ This is true because $\sqrt{kx}J_{n}\left(  kx\right)  $ are
energy eigenfunctions of $H_{0}$ for all real $k>0$. \ Explicitly
\cite{Radial},%
\begin{equation}
\frac{\hbar^{2}}{2m}\left(  -\frac{d^{2}}{dx^{2}}+\frac{n^{2}-1/4}{x^{2}%
}\right)  \left(  \sqrt{kx}J_{n}\left(  kx\right)  \right)  =\frac{\hbar
^{2}k^{2}}{2m}~\sqrt{kx}J_{n}\left(  kx\right)
\end{equation}
Orthogonality and completeness of these eigenfunctions are now expressed as a
superfluous pair of equations%
\begin{equation}
\int_{0}^{\infty}\sqrt{k_{1}x}J_{n}\left(  k_{1}x\right)  \sqrt{k_{2}x}%
J_{n}\left(  k_{2}x\right)  dx=\delta\left(  k_{1}-k_{2}\right)
\ ,\ \ \ \int_{0}^{\infty}\sqrt{kx_{1}}J_{n}\left(  kx_{1}\right)
\sqrt{kx_{2}}J_{n}\left(  kx_{2}\right)  dk=\delta\left(  x_{1}-x_{2}\right)
\end{equation}
That is to say, for the system with a repulsive $1/x^{2}$ potential, the
functions $\sqrt{kx}J_{n}\left(  kx\right)  $ play a role similar to the plane
waves for the free particle. \ 

The propagator for $H_{0}$ is therefore given by an integral involving Bessel
function bilinears, namely,
\begin{equation}
\left\langle x_{1}\right\vert \exp\left(  -iH_{0}t/\hbar\right)  \left\vert
x_{2}\right\rangle =\int_{0}^{\infty}\sqrt{kx_{1}}J_{n}\left(  kx_{1}\right)
e^{-\frac{i\hbar k^{2}t}{2m}}\sqrt{kx_{2}}J_{n}\left(  kx_{2}\right)  dk
\end{equation}
It so happens this integral reduces to a closed form in terms of another,
modified Bessel function, $I_{n}$ (e.g. see \cite{Peak}). \
\begin{equation}
\int_{0}^{\infty}\sqrt{kx_{1}}J_{n}\left(  kx_{1}\right)  e^{-\frac{i\hbar
k^{2}t}{2m}}\sqrt{kx_{2}}J_{n}\left(  kx_{2}\right)  dk=\frac{m\sqrt
{x_{1}x_{2}}}{i\hbar t}~I_{n}\left(  \frac{mx_{1}x_{2}}{i\hbar t}\right)
\exp\left(  \frac{im}{2\hbar t}\left(  x_{1}^{2}+x_{2}^{2}\right)  \right)
\end{equation}
Thus the result for $H_{0}$, at least when the potential is repulsive, is
\begin{equation}
\left\langle x_{1}\right\vert \exp\left(  -iH_{0}t/\hbar\right)  \left\vert
x_{2}\right\rangle =\frac{m\sqrt{x_{1}x_{2}}}{i\hbar t}~I_{n}\left(
\frac{mx_{1}x_{2}}{i\hbar t}\right)  \exp\left(  \frac{im}{2\hbar t}\left(
x_{1}^{2}+x_{2}^{2}\right)  \right)
\end{equation}
Combining this result with (\ref{Element})\ then gives the propagator for
$\left.  H\left(  \lambda,\omega\right)  \right\vert _{\lambda=\hbar
^{2}\left(  n^{2}-1/4\right)  }$ upon re-parameterizing the time,
$t\rightarrow\frac{\sin\left(  \omega t\right)  }{\omega}$, in agreement with
long-known results (again see \cite{Peak}, as well as Section 3.3 in
\cite{Handbook}). \ \
\begin{align}
&  \left\langle x_{1}\right\vert \exp\left(  -iHt/\hbar\right)  \left\vert
x_{2}\right\rangle \nonumber\\
&  =\frac{m\omega\sqrt{x_{1}x_{2}}}{i\hbar\sin\left(  \omega t\right)  }%
~I_{n}\left(  \frac{m\omega x_{1}x_{2}}{i\hbar\sin\left(  \omega t\right)
}\right)  \exp\left(  -\frac{im\omega}{2\hbar}~x_{1}^{2}\tan\left(  \omega
t/2\right)  +\frac{im\omega}{2\hbar\sin\left(  \omega t\right)  }\left(
x_{1}^{2}+x_{2}^{2}\right)  -\frac{im\omega}{2\hbar}~x_{2}^{2}\tan\left(
\omega t/2\right)  \right) \nonumber\\
&  =\frac{m\omega\sqrt{x_{1}x_{2}}}{i\hbar\sin\left(  \omega t\right)  }%
~I_{n}\left(  \frac{m\omega x_{1}x_{2}}{i\hbar\sin\left(  \omega t\right)
}\right)  \exp\left(  \frac{im\omega}{2\hbar}~\left(  x_{1}^{2}+x_{2}%
^{2}\right)  \cot\left(  \omega t\right)  \right)  \label{HalflineSHO}%
\end{align}
where $\cot\left(  2\theta\right)  =-\tan\left(  \theta\right)  +1/\sin\left(
2\theta\right)  $ was once again used in the last step.

It is perhaps reassuring that for short times, and both $x_{1},x_{2}>0$,
\begin{equation}
\left\langle x_{1}\right\vert \exp\left(  -iHt/\hbar\right)  \left\vert
x_{2}\right\rangle \underset{t\rightarrow0}{\sim}\sqrt{\frac{m}{2\pi i\hbar
t}}\exp\left(  \frac{im}{2\hbar t}\left(  x_{1}-x_{2}\right)  ^{2}\right)
\end{equation}
as follows from the asymptotic behavior of $I_{n}$. \ Consequently,
$\lim_{t\rightarrow0}\left\langle x_{1}\right\vert \exp\left(  -iHt/\hbar
\right)  \left\vert x_{2}\right\rangle =\delta\left(  x_{1}-x_{2}\right)  $.
\ Even more so, it is straightforward to check that%
\begin{equation}
\frac{\hbar^{2}}{2m}\left(  -\frac{\partial^{2}}{\partial x_{1}^{2}}%
+\frac{n^{2}-1/4}{x_{1}^{2}}+\frac{m^{2}\omega^{2}x_{1}^{2}}{\hbar^{2}%
}\right)  \left\langle x_{1}\right\vert \exp\left(  -iHt/\hbar\right)
\left\vert x_{2}\right\rangle =i\hbar\frac{\partial}{\partial t}\left\langle
x_{1}\right\vert \exp\left(  -iHt/\hbar\right)  \left\vert x_{2}\right\rangle
\end{equation}
These properties guarantee the time evolution of a wave function $\psi\left(
x,t\right)  $ defined on the half-line is correctly given by $\psi\left(
x_{1},t\right)  =\int_{0}^{\infty}\left\langle x_{1}\right\vert \exp\left(
-iHt/\hbar\right)  \left\vert x_{2}\right\rangle \psi\left(  x_{2},0\right)
dx_{2}$ \cite{Phases}.

\subsubsection*{Appendix: \ Other Identities}

There are many other $SL\left(  2,%
\mathbb{C}
\right)  $ \textquotedblleft roads\textquotedblright\ that lead to a closed
form expression for the oscillator propagator (i.e. the \textquotedblleft
Rome\textquotedblright\ of this problem).\ \ Here we list a few more of them
involving the operators $x$ and $p$ with $\left[  x,p\right]  =i\hbar$. \ For
simplicity, we omit the $\lambda/x^{2}$\ potential, but the identities to
follow are valid even with that potential term upon substituting
$p^{2}\rightarrow p^{2}+\lambda/x^{2}$.

In addition to the identity in the main text, (\ref{TSvKIdentity}), there are
many relations paired by conjugation plus $t\rightarrow-t$. \ For example,%
\begin{gather}
\exp\left(  -\frac{it}{\hbar}\left(  \frac{1}{2m}p^{2}+\frac{1}{2}m\omega
^{2}x^{2}\right)  \right)  =\exp\left(  -i\alpha x^{2}\right)  \exp\left(
-i\gamma\left(  xp+px\right)  \right)  \exp\left(  -i\beta p^{2}\right)
\tag{A1a}\\
\text{with \ \ }e^{2\gamma\hbar}=\cos\left(  \omega t\right)  \ ,\ \ \ \alpha
=\frac{m\omega}{2\hbar}\tan\left(  \omega t\right)  \ ,\ \ \ \beta=\frac
{1}{2\hbar m\omega}\tan\left(  \omega t\right) \nonumber
\end{gather}%
\begin{gather}
\exp\left(  -\frac{it}{\hbar}\left(  \frac{1}{2m}p^{2}+\frac{1}{2}m\omega
^{2}x^{2}\right)  \right)  =\exp\left(  -i\beta p^{2}\right)  \exp\left(
-i\gamma\left(  xp+px\right)  \right)  \exp\left(  -i\alpha x^{2}\right)
\tag{A1b}\\
\text{with \ \ }e^{-2\gamma\hbar}=\cos\left(  \omega t\right)  \ ,\ \ \ \alpha
=\frac{m\omega}{2\hbar}\tan\left(  \omega t\right)  \ ,\ \ \ \beta=\frac
{1}{2\hbar m\omega}\tan\left(  \omega t\right) \nonumber
\end{gather}%
\begin{gather}
\exp\left(  -\frac{it}{\hbar}\left(  \frac{1}{2m}p^{2}+\frac{1}{2}m\omega
^{2}x^{2}\right)  \right)  =\exp\left(  -i\gamma\left(  xp+px\right)  \right)
\exp\left(  -i\alpha x^{2}\right)  \exp\left(  -i\beta p^{2}\right)
\tag{A2a}\\
\text{with \ \ }e^{2\gamma\hbar}=\cos\left(  \omega t\right)  \ ,\ \ \ \alpha
=\frac{m\omega}{2\hbar}\sin\left(  \omega t\right)  \cos\left(  \omega
t\right)  \ ,\ \ \ \beta=\frac{1}{2\hbar m\omega}\tan\left(  \omega t\right)
\nonumber
\end{gather}%
\begin{gather}
\exp\left(  -\frac{it}{\hbar}\left(  \frac{1}{2m}p^{2}+\frac{1}{2}m\omega
^{2}x^{2}\right)  \right)  =\exp\left(  -i\beta p^{2}\right)  \exp\left(
-i\alpha x^{2}\right)  \exp\left(  -i\gamma\left(  xp+px\right)  \right)
\tag{A2b}\\
\text{with \ \ }e^{-2\gamma\hbar}=\cos\left(  \omega t\right)  \ ,\ \ \ \alpha
=\frac{m\omega}{2\hbar}\sin\left(  \omega t\right)  \cos\left(  \omega
t\right)  \ ,\ \ \ \beta=\frac{1}{2\hbar m\omega}\tan\left(  \omega t\right)
\nonumber
\end{gather}%
\begin{gather}
\exp\left(  -\frac{it}{\hbar}\left(  \frac{1}{2m}p^{2}+\frac{1}{2}m\omega
^{2}x^{2}\right)  \right)  =\exp\left(  -i\alpha x^{2}\right)  \exp\left(
-i\beta p^{2}\right)  \exp\left(  -i\gamma\left(  xp+px\right)  \right)
\tag{A3a}\\
\text{with \ \ }e^{2\gamma\hbar}=\cos\left(  \omega t\right)  \ ,\ \ \ \alpha
=\frac{m\omega}{2\hbar}\tan\left(  \omega t\right)  \ ,\ \ \ \beta=\frac
{\sin\left(  \omega t\right)  \cos\left(  \omega t\right)  }{2\hbar m\omega
}\nonumber
\end{gather}%
\begin{gather}
\exp\left(  -\frac{it}{\hbar}\left(  \frac{1}{2m}p^{2}+\frac{1}{2}m\omega
^{2}x^{2}\right)  \right)  =\exp\left(  -i\gamma\left(  xp+px\right)  \right)
\exp\left(  -i\beta p^{2}\right)  \exp\left(  -i\alpha x^{2}\right)
\tag{A3b}\\
\text{with \ \ }e^{-2\gamma\hbar}=\cos\left(  \omega t\right)  \ ,\ \ \ \alpha
=\frac{m\omega}{2\hbar}\tan\left(  \omega t\right)  \ ,\ \ \ \beta=\frac
{\sin\left(  \omega t\right)  \cos\left(  \omega t\right)  }{2\hbar m\omega
}\nonumber
\end{gather}
These identities immediately lead to the propagator (\ref{SHO}), or
(\ref{HalflineSHO}) after restoration of the $\lambda/x^{2}$ term, upon taking
into account the effects of $\exp\left(  -i\gamma\left(  xp+px\right)
\right)  $ to rescale position eigenstates. \ Namely,%
\begin{equation}
\exp\left(  -i\gamma\left(  xp+px\right)  \right)  ~\left\vert x_{2}%
\right\rangle =\exp\left(  \hbar\gamma\right)  ~\left\vert x_{2}\exp\left(
2\hbar\gamma\right)  \right\rangle \tag{A4}%
\end{equation}
etc.


\begin{thebibliography}{99}
\bibitem {Exercise}As another exercise, the reader is invited to find
parameter-dependent matrix realizations of the $SL\left(  2,%
\mathbb{C}
\right)  $ group elements and use them to establish the identity
(\ref{TSvKIdentity}). \ Hint: \ The $sl\left(  2,%
\mathbb{C}
\right)  $ algebra (\ref{sl2}) is also realized by
\[
x^{2}\rightarrow\left(
\begin{array}
[c]{cc}%
0 & 2\hbar/\lambda\\
0 & 0
\end{array}
\right)  \ ,\ \ \ p^{2}+\frac{\lambda}{x^{2}}\rightarrow\left(
\begin{array}
[c]{cc}%
0 & 0\\
2\hbar\lambda & 0
\end{array}
\right)  \ ,\ \ \ xp+px\rightarrow\left(
\begin{array}
[c]{cc}%
-2i\hbar & 0\\
0 & 2i\hbar
\end{array}
\right)
\]
for any $\lambda$. \ One may now use copies of this realization for various
different $\lambda$, including the $\lambda=1$ matrices of (\ref{2x2}), to
build larger matrix realizations of the algebra and the group elements.
\end{thebibliography}
\end{document}